\documentclass{PoS}
\newcommand{\be}{\begin{equation}}
\newcommand{\ee}{\end{equation}}
\usepackage{amsmath}

\title{On the cosmic ray spectrum from type II Supernovae expanding in their red giant presupernova wind}

\ShortTitle{NRH instability and type II Supernovae}

\author{\speaker{Martina Cardillo}\\
        INAF-Osservatorio Astrofisico di Arcetri\\
        E-mail: \email{martina@arcetri.astro.it}}

\author{Elena Amato\\
        INAF-Osservatorio Astrofisico di Arcetri\\
        E-mail: \email{amato@arcetri.astro.it}}
        
\author{Pasquale Blasi\\
        INAF-Osservatorio Astrofisico di Arcetri $\&$ Gran Sasso Science Institute (INFN)\\
        E-mail: \email{blasi@arcetri.astro.it}}

\abstract{One of the most important challenges for the largely accepted idea that Galactic CRs are accelerated in SNR shocks
is the maximum energy at which particles can be accelerated. The resonant streaming instability, long invoked for magnetic field amplification at shocks, can not provide sufficiently high fields and efficient enough scattering so as to ensure particle acceleration up to the \textit{knee}. Here we discuss the non-resonant version of this instability which, with its faster growth and larger value of the amplified field, increases the achievable maximum energy. Because of their higher explosion rate, we focus on type II SNe expanding in their red supergiant wind and we find that the transition between Ejecta Dominated (ED) and Sedov-Taylor (ST) phases takes place at very early times. In this environment, the accelerated particle spectrum shows no high energy exponential cut-off but a spectral break at the maximum energy (E$_{M}$). Moreover, the maximum energy of protons can easily reach PeV energies. With this model, we tried to fit KASCADE-Grande and ARGO -YBJ data but failed to find a parameter combination that can explain both data sets. We discuss the different scenarios implied by the two data sets.} 

\FullConference{The 34th International Cosmic Ray Conference,\\
		30 July- 6 August, 2015\\
		The Hague, The Netherlands}

\begin{document}

\section{Introduction}
\label{intro}
Supernova remnants (SNRs) are considered the most probable sources of Galactic CRs based on energetic argument s, Supporting evidence has been found in the X-ray band where amplified magnetic fields have been inferred at the shock with surrounding medium (for reviews see, e.g., \cite{blasi13rev,amato14rev}). Magnetic field amplification, not only downstream but also upstream of the shock, is fundamental in order to obtain high energies. The quest for understanding what kind of instability could lead to the right conditions is still open. The resonant instability induced by CR streaming leads to estimate $E_{M}\sim 10^{3}-10^{4}$ GeV \cite{lagage83a,lagage83b}, still well below the \textit{knee}: this is due to its resonant nature that fixes the saturation at $\delta B/B\sim 1$ \cite{kulsrud69,wentzel74,skilling75a,skilling75b,skilling75c}. Non-resonant modes, largely studied in Refs.~\cite{lucek00} and \cite{bell04}, starting with a wavenumber $k$ much larger than the particle gyroradius but then forming fluctuations on larger scales allowing resonant scattering, grow in a very short time. This kind of instability may allow to reach the \textit{knee} if applied to the case of supernovae expanding in the wind of their red super-giant (RSG) progenitor star \cite{bell13,schure13}, where the highest energies are reached within a few decades after the SN explosion, namely before the beginning of the ST phase of the explosion.\\ 
The origin of the \textit{knee} of the all-particle spectrum is a very challenging and a very important issue because it is strongly correlated with the transition between galactic and extragalactic CRs. In Ref.~\cite{aloisio14}, in order to fit the Auger data \cite{auger10,HiREs04,HiREs08,TA11}, that require a predominant light component at EeV energies, the authors introduce an extra component of protons; this added proton population is in good agreement with the proton spectrum measured by KASCADE-Grande \cite{apel13} (see also the results of ICETOP \cite{ICETOP13}). However, proton and iron spectra detected by KASCADE-Grande in the energy region $10^{16}-10^{18}$ eV make the issue of understanding what the maximum achievable energy is even more challenging: these data can be explained only with a CR spectrum that is not cut-off exponentially at the maximum energy, or with a new unknown, class of sources with maximum energy much in excess of the \textit{knee}.  Furthermore, recently, ARGO-YBJ \cite{argo} and YAC1-Tibet Array \cite{yac} experiments measured a \textit{knee} in the light component at $\sim 650$ TeV, appreciably below the PeV energy of the all-particle \textit{knee}, making the situation ever more complex. 

In this work we analyse the probability of reaching the energy of the 'knee' in particle acceleration during the expansion of a SN shock in the dense region occupied by the wind of a RSG. In particular, we investigated in detail the implications of the so called Non-Resonant Hybrid (NRH) instability described by Refs.~\cite{bell04,schure13,bell13} by computing the maximum energy and the overall particle spectrum produced during the whole SN expansion, both in the ED and ST phases, and comparing it with KASCADE-Grande and ARGO data \cite{cardillo15}.

\section{Model}
\label{sec-2}
\subsection{Maximum Energy}
\label{sub-2.1}
The non-resonant instability has a very high growth rate and, despite the fastest mode has a wavelength much shorter than the Larmor radius of the particles generating the current, an 'inverse cascade' mechanism during its non-linear evolution, increases the spatial scale involved \cite{bell04, riquelme09,caprioli14}.
Particles with energy equal to the current achievable maximum, $E_{M}$, move from downstream to upstream where, by definition, they can not  be scattered resonantly  and escape the system moving quasi-balistically at a speed close to that of light. In order to obtain an estimation of the maximum energy achievable, we passed through different steps.
First of all, we write the current, at distance $R$ from the explosion center, due to particles escaped when the shock location was at $r$:
\begin{equation}
j_{CR}(R,r)=n_{\rm CR,r}\left(R,E_M(r)\right) e v_{sh}(r) =e \frac{\xi_{CR} \rho(r) v_s^3}{E_0\Psi(E_M)}\left(\frac{r}{R}\right)^2
\label{eq:jCR}
\end{equation}
where $\xi_{CR}$ is the CR acceleration efficiency and
\begin{equation}
\Psi=\left\{
\begin{array}{ll}
\left(E_M/E_0\right)\ln \left(E_M/E_0\right) & \beta=0\\
 & \\
\frac{1+\beta}{\beta}\left(\frac{E_M}{E_0}\right)^{1+\beta}\left[1-\left(\frac{E_0}{E_M}\right)^\beta\right] & \beta\ne0\ .
\end{array}
\right.
\label{eq:psi}
\end{equation}
The function $\Psi(E_M)$ accounts for normalisation of the particle distribution function which is taken to be $f_s(E)\propto E^{-(2+\beta)}$ and extending between a minimum energy $E_0$, that does not depend on time and a maximum energy $E_M$ which does depend on time, as we will see (details in \cite{cardillo15}).
Numerical simulations show that the saturation of the instability occurs after $N_t\sim 5$ e-folds \cite{bell04}, namely 
\begin{equation}
\int \gamma_{M}dt=\int (k_{M}v_{A}) dt \approx 5\ .
\label{eq:growthrate}
\end{equation}
where $v_{A}$ is the Alfv\'en speed in the unperturbed magnetic field $B_{0}$.
The wavenumber of the fastest growing modes, $k_{M}$, can be easily estimated considering balance between current and magnetic tension:
\begin{equation}
k_{M} B_{0}\cong\frac{4\pi}{c}j_{CR},
\label{eq:Btension}
\end{equation}
We can see that it depends on CR current (Eq.~\ref{eq:jCR}) and, consequently, on the density profile. For a supernova expanding in the uniform ISM, the density can be assumed constant $\rho(R)=\rho_{ISM}$. For a type II supernova expanding its pre-SN wind, instead, the density profile can be written as $\rho(R)\cong\frac{\dot{M}}{4\pi R_{0}^{2}V_{w}}\left(\frac{R_0}{R}\right)^{2}=\rho(R_{0})\left(\frac{R_{0}}{R}\right)^{2}$, where $\dot{M}$ is the rate of mass loss of the red giant and $V_{w}$ is the wind velocity. In the following we will simply write $\rho(R)\propto R^{-m}$ and consider the two cases: $m=0$, that corresponds to a uniform medium, and $m=2$, that describes expansion in the progenitor wind.
At this point, differentiating Eq.~\ref{eq:growthrate} with respect to $R$, we obtain an implicit expression for the maximum energy:
\begin{equation}
\Psi(E_{M}(R))\cong\frac{2 e}{(4-m)5c E_0}\xi_{CR} v_s(R)^2 \sqrt{4 \pi \rho(R) R^2}\ .
\label{eq:Em_general}
\end{equation}
From this equation, we can find maximum energy for type I and type II SNae for different injection spectral indices. 
In the case of a $E^{-2}$ ($\beta=0$) source spectrum, for type I events we can estimate:
\be
\begin{split}
E_{M}&\cong\frac{2 e}{10c}\xi_{CR} v_{0}^2 \sqrt{4 \pi \rho R_{0}^2}=\\
&130\ \left(\frac{\xi_{CR}}{0.1}\right) \left(\frac{M_{ej}}{M_\odot}\right)^{-\frac{2}{3}} \left(\frac{E_{SN}}{10^{51}\rm{erg}}\right) \left(\frac{n_{ISM}}{\rm{cm^{-3}}}\right)^{\frac{1}{6}}\ TeV
\label{eq:emaxI}
\end{split}
\ee
while for type II:
\be
\begin{split}
&E_{M}\cong\frac{2 e}{5c}\xi_{CR} v_{0}^2 \sqrt{4 \pi \rho R_{0}^2}\approx\\
&1\ \left(\frac{\xi_{CR}}{0.1}\right)\left(\frac{M_{ej}}{M_\odot}\right)^{-1} \left(\frac{E_{SN}}{10^{51}\rm{erg}}\right) \left(\frac{\dot M}{10^{-5} M_\odot /yr}\right)^{\frac{1}{2}} \left(\frac{V_w}{10\,\rm{km/s}}\right)^{-\frac{1}{2}}\ PeV
\label{eq:emaxII}
\end{split}
\ee
It is apparent that for standard parameters pertaining the two types of explosions, type II SNe can reach about one order of magnitude larger maximum energies than type Ia, and in particular that if particles are accelerated and escape as described here, these sources easily seem to be able to provide proton acceleration up to the \textit{knee} (details in \cite{cardillo15}) .

\subsection{Particle spectrum}
\label{sub:2.2}
In order to obtain the particle spectral behavior provided by NRH instability, we use the expression in Eq.~\ref{eq:jCR} for the current and assuming a general power-law dependence of the shock radius on time: $R\propto t^\lambda$, which also implies $v_s\propto R^{(\lambda-1)/\lambda}$.  So, the escaping particle spectrum is:
\be
N_{esc}(E)=\frac{J_{CR}}{e} 4 \pi R^2 \frac{dt}{dE}=\frac{4 \pi \xi_{CR}}{E_0 S(\lambda,m)}\rho v_s^2 R^3\chi(E)\ ,
\label{eq:nescfin}
\ee
where $\chi(E)=(d/dE)(1/\Psi(E))$ is a function of $E$ alone, to be computed from Eq.~\ref{eq:psi}  and $S(\lambda,m)$ depends on the details of the SNR evolution (more detailed explanation can be found in \cite{cardillo15}). 
When computing $\chi(E)$ explicitly from Eq.~\ref{eq:psi} we find:
\be
\chi(E)\approx\frac{d}{dE}\left(\frac{1}{\Psi}\right)=\frac{1}{E_0}
\left\{
\begin{array}{ll}
\frac{\beta}{1+\beta} \left(\frac{E}{E_0}\right)^{-2} & \beta<0\\
&   \\
\left(\frac{E}{E_0}\right)^{-2}\left[\frac{\left(1+\ln(E/E_0)\right)}{(\ln(E/E_0))^{2}}\right] & \beta=0 \\
 & \\
\beta \left(\frac{E}{E_0}\right)^{-(2+\beta)} & \beta>0
\end{array}
\right.
\label{eq:dpside}
\ee
where the sign $\approx$ refers to the fact that we have used the assumption $E>>E_0$. It is apparent that the power-law dependence of $\chi(E)$ can only be $-2$ or steeper. Therefore the spectrum of CRs released during the Sedov-Taylor phase is the same as the spectrum of accelerated particles in the source only if the latter is $E^{-2}$ or steeper, while for flatter source spectra $N_{esc}(E)\propto E^{-2}$. 

In the following, we take the value of $\lambda$ appropriate to describe the different evolutionary stages of the SNR from Ref.~\cite{chevalier82,chevalier89,ptuskin05,truelove99}. These works give the remnant expansion law during the different stages for generic density profiles of both the SN ejecta $\rho_{ej}\propto R^{-k}$ and the ambient medium $\rho\propto R^{-m}$. 
Once $\lambda$ is given, from Eq.~\ref{eq:nescfin} and \ref{eq:dpside} we can derive the spectrum released by the SN during the ED and ST phases as \cite{cardillo15}:
\be
N_{esc}(E)\propto 
\left\{ 
\begin{array}{ll}
E^{-(5+4\epsilon)}& \,\,\,\,\,\textrm{ED phase;}\\
 & \, \, \, \, \, \, \, \,\, \, \, \,\, \, \, \,\, \, \, \,\, \, \, \,m=0, k=7\\
E^{-(2+\epsilon)}& \,\,\,\,\,\textrm{ST phase;}
\end{array} 
\right.
\ee
\be
N_{esc}(E)\propto\left\{ 
\begin{array}{ll}
E^{-(4+3\epsilon)}& \,\,\,\,\,\textrm{ED phase;}\\
 & \, \, \, \,\, \, \, \,\, \, \, \,\, \, \, \,\, \, \, \,\, \, \, \,m=2, k=9\\
E^{-(2+\epsilon)}& \,\,\,\,\,\textrm{ST phase;}
\end{array} \right.
\label{eq:spec}
\ee
This result is very interesting because it shows that the spectrum of escaping CRs integrated over time during the SN shock expansion is a broken power law, with an index close to 2 at low energies (below$E_{M}$) and steeper at higher energies. There is no trace of the exponential cut-off expected by standard DSA theory. The transition between the two power laws at the maximum energy $E_{M}$ corresponds to the transition between the end of the ED phase, where most mass has already been processed, and the beginning  of the ST (adiabatic) phase.

In the next step we take into account the transport of nuclei in the Galaxy. Assuming that the transport has reached a stationary regime and that it is dominated by diffusion and  that both injection and spallation occur in a thin disc of size $2h$, the solution of the transport equation in the Galactic disc can be easily written as follows:
\begin{equation}
f_{0,\alpha}(p) = \frac{N_{esc}(p) \Re}{\pi R_{D}^{2} \mu v} \frac{X(p)}{1 + \frac{X(p)}{X_{\alpha}}},
\end{equation}
where we introduced the grammage $X(p) = n_{d} \frac{h}{H} v m_{p} \frac{H^{2}}{D(p)}$. Here $n_{d}$ is the gas density of the Galactic disc, assumed of thickness $2h$ and radius $R_{D}$, and the escaping spectrum $N_{esc}(p)$ is calculated as discussed above but using a parametric form for $R_{sh}$ in order to model the transition from the ED to the ST phase \cite{cardillo15}. Written in this form, the solution is very clear. At particle momenta for which the grammage is small compared with $X_{\alpha}$ the standard solution is recovered $f_{0,\alpha}(p) = \frac{N_{\alpha}(p) \Re}{2 \pi R_{D}^{2} H} \frac{H^{2}}{D(p)}$ while in the range of momenta at which the solution is spallation-dominated, the observed spectrum reproduces the shape of the injection spectrum. 

In order to calculate the spectra of CRs at the Earth, an evaluation of the grammage is needed and we use the formulation of Ref.~\cite{Ptuskin09} as a starting point. Given that the CR spectrum observed at Earth has an energy dependence $E^{- 2.65}$ as derived from TRACER and CREAM data \cite{ave08,yoon11}, in our calculations we used a slope of the diffusion coefficient $\delta=2.65-p_{inj}$, where for $p_{inj}$ we considered the values 2 and 2.31. In the first case, the diffusion coefficient has a slightly stronger energy dependence than the one used by \cite{Ptuskin09} ($E^{0.6}$); in the other case, instead, we use the exact form given in that work. For the spallation cross section we adopt the simple formulation of Ref.~\cite{horandel07} (details in \cite{cardillo15}).

\section{Results}
\label{sec-3}
\begin{figure}[!h]
\centering
\includegraphics[scale=.7]{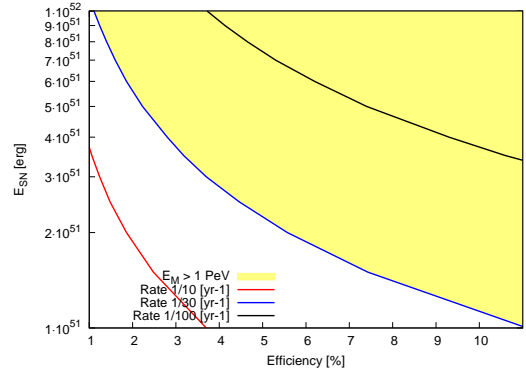}
\caption{ Relations between physical parameters in the case of injection index $p=-2$. The shaded area in the $\xi_{CR}-E_{SN}$ plane indicates the allowed range of parameters to reach $E_{M}=1$ PeV. The different curves indicate the combination of values of the parameters for which the observed proton flux is also fitted. Each line refers to a given SN rate (as labeled).}
\label{fig:parameters}
\end{figure}

Type II SNe are the most likely sources to accelerate protons up to the \textit{knee}. We focus on this type of explosions in order to find a fit for the all particle spectrum. With this aim we vary the different parameters, considering the right ejecta density profile.
We have seen that the value of $E_{M}$ is determined by the combination of the CR acceleration efficiency, $\xi_{CR}$, and the energetics of the SN, $E_{SN}$. On the other hand, the flux of CRs at the Earth derives from a combination of $\xi_{CR}$, $E_{SN}$ and the rate of SNe, $\Re$. Fitting the proton spectrum in the \textit{knee} region imposes an additional constraint on the combination $\xi_{CR}$, $E_{SN}$, $\Re$ (see Fig.~\ref{fig:parameters}). 

We calculated the required efficiency and rates necessary to reach the \textit{knee} and fit the overall CR flux: for steep spectra at the shock, the escape current is lower and this forces one to have larger SN energetics and larger CR acceleration efficiencies, which is counterintuitive for a steep spectrum of accelerated particles. For flat spectra at the shock ($E^{-2}$), the spectrum of escaping particles is also flat and PeV energies can be reached for ordinary parameters of the SN explosion. Reaching PeV energies is even easier for hard spectra at the shock. Our calculations show that a type II SN with standard energetics ($E_{SN}=10^{51}$ erg, $\xi_{CR}=10\%$) can accelerate CRs up to energy of the \textit{knee}, $E_{M}\sim1$ PeV. The spectrum obtained with these parameters is shown in Fig.~3 of \cite{cardillo15}. 

Expressing the time dependence of the maximum energy through the parametrization used in this work, we can show that PeV energies can be reached at very early time, $\sim 10-30$ years after the SN explosion, . This result not only changes the SNR paradigm according to which the highest energies are reached  several hundred years after explosion, but also narrows down the probability of catching a PeVatron in action in our own Galaxy.

\begin{figure}[!ht]
\centering
\includegraphics[scale=0.7]{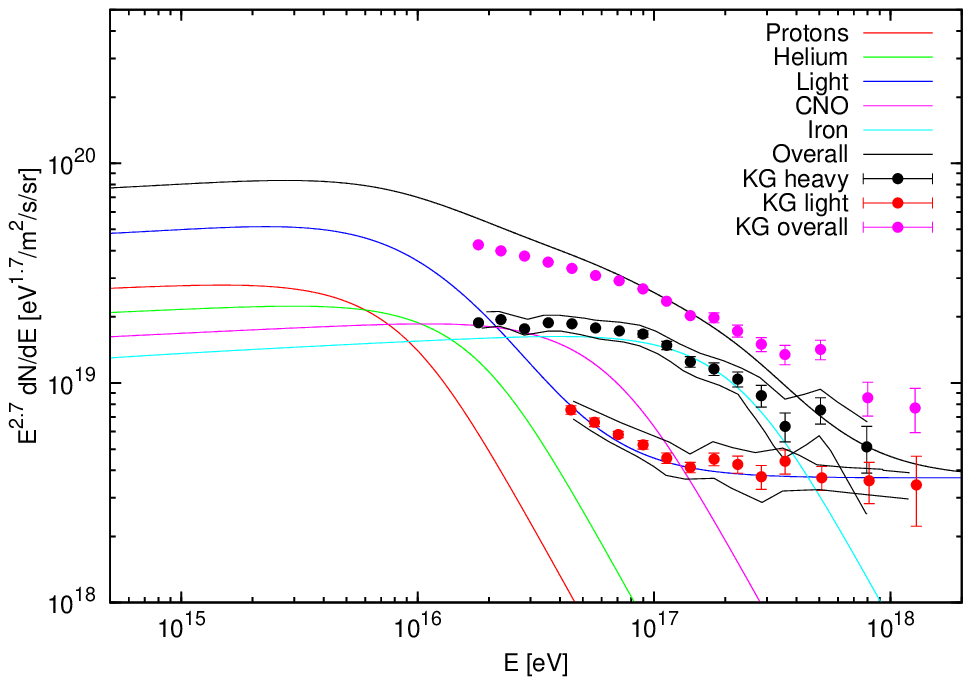}  
\includegraphics[scale=0.7]{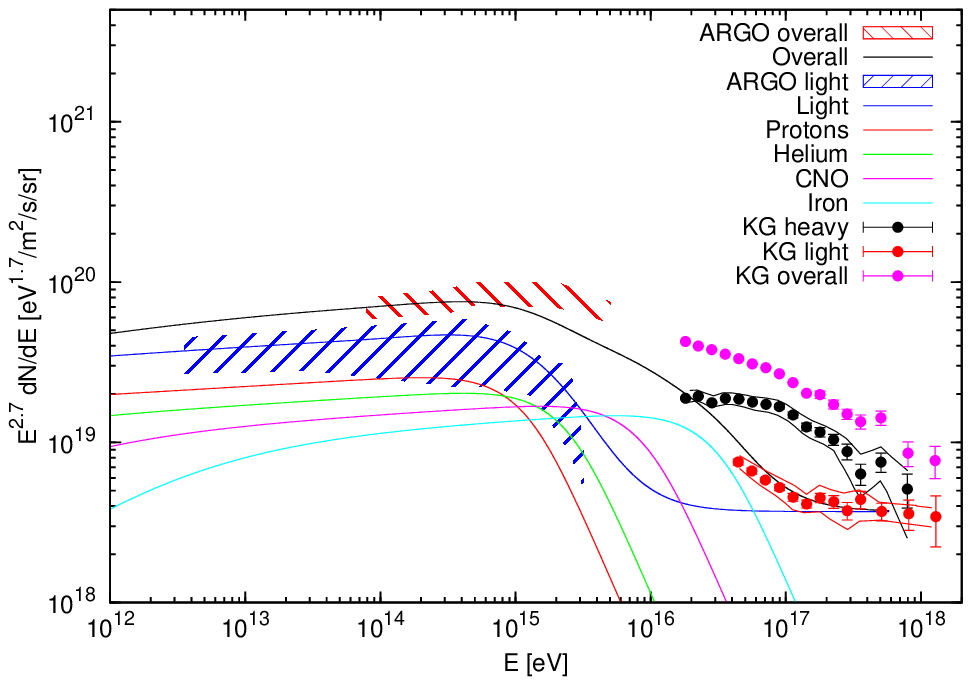} 
\caption{{(\it Left)} our best fit model for KASCADE-Grande data \cite{apel13}, with $k=9$, $E_{SN}=2\times10^{51}$ erg, $\Re=1/110~yr^{-1}$, $E_{M}^{H}\cong 3.7$ PeV and $\xi_{CR}\cong20\%$. The highest energy data are fitted with an "ad hoc" extragalactic component. {(\it Right)} Best fit model to ARGO data \cite{argo,disciascio14}, with $k=9$, $E_{SN}=10^{51}$ erg, $\Re=1/15~yr^{-1}$, $E_{M}\cong 507$ TeV and $\xi_{CR}\cong5.2\%$.}
\label{fig:dataKG_ARGO}
\end{figure}

Finally, we used our model for an attempt to fit recent challenging data of KASCADE-Grande and ARGO. 
We can fit individually both sets of data, using reasonable values of the SN parameters and CR acceleration efficiencies; by the way, there is no model that can fit both at the same time (see Fig.~\ref{fig:dataKG_ARGO}). KASCADE-Grande data require a SN energy $E_{SN}=2\times10^{51}$erg, an efficiency $\xi_{CR}=20\%$ and a rate of explosion of $\Re=1/110~\rm{yr^{-1}}$ that lead to a maximum energy $E_{M}\cong 3.7$ PeV, whereas we can fit ARGO data with $E_{SN}=10^{51}$ erg, $\Re=1/15~yr^{-1}$ and $\xi_{CR}\cong5.2\%$ that lead to $E_{M}\cong 507$ TeV.

The different results obtained by these two experiments measuring the same quantities points to the possible underestimate of systematic uncertainties involved in the adopted experimental techniques.


\begin{thebibliography}{}

\bibitem{ICETOP13} M.G.Aartsen et al., \textit{Measurement of the cosmic ray energy spectrum with IceTop-73},\textit{Phys. Rev.} \textbf{D 88} 042004 (2013) 
%
\bibitem{HiREs04} R. Abbasi et. al., \textit{Measurement of the Flux of Ultrahigh Energy Cosmic Rays from Monocular Observations by the High Resolution Fly's Eye Experiment},\textit{Phys. Rev. Lett.} \textbf{92}, 151101 (2004) [astro-ph/0208243]
%
\bibitem{HiREs08} R. Abbasi et. al., \textit{First Observation of the Greisen-Zatsepin-Kuzmin Suppression},\textit{Phys.Rev.Lett.},\textbf{100} (2008) 101101, [arXiv:astro-ph/0703099] 
%
\bibitem{auger10} J. Abraham et. al., \textit{Measurement of the Depth of Maximum of Extensive Air Showers above 1018eV}, \textit{Phys.Rev.Lett.} \textbf{104} (2010) 091101 [arXiv:1002.0699]
%
\bibitem{adriani11} O. Adriani et al., \textit{PAMELA Measurements of Cosmic-Ray Proton and Helium Spectra}, \textit{Science} \textbf{332} (2011) 69 [arXiv:1103.4055]
%
\bibitem{aloisio14} R. Aloisio et al., \textit{Ultra high energy cosmic rays: implications of Auger data for source spectra and chemical composition}, \textit{JCAP} \textbf{10} (2014) 020A [arXiv1312.7459]
%
\bibitem{amato14rev} E. Amato, \textit{The origin of galactic cosmic rays}, \textit{JMPD} \textbf{23} (2014) 1430013 [arXiv:1406.7714]
%
\bibitem{apel13} W. Apel et. al., \textit{Ankle-like feature in the energy spectrum of light elements of cosmic rays observed with KASCADE-Grande}, \textit{Phys. Rev.} \textbf{87} (2013) 1101 [arXiv:1304.7114]
%
\bibitem{ave08} M. Ave et al., \textit{Composition of Primary Cosmic-Ray Nuclei at High Energies}, \textit{ApJ},\textbf{678} (2008) 262 [arXiv:0801.0582]
%
\bibitem{bell04} A.R. Bell, \textit{Turbulent amplification of magnetic field and diffusive shock acceleration of cosmic rays}, \textit{MNRAS} \textbf{353} (2004) 550-558 
%
\bibitem{bell13}A.R. Bell et al., \textit{Cosmic-ray acceleration and escape from supernova remnants}, \textit{MNRAS} \textbf{431} (2013) 415-429 [arXiv:1301.7264]
%
\bibitem{blasi13rev} P. Blasi, \textit{The origin of galactic cosmic rays}, \textit{A$\&$ARv} \textbf{21} (2013) 70 [arXiv:1311.7346]
%
\bibitem{blasi14} P. Blasi, \textit{Origin of very high- and ultra-high-energy cosmic rays}, \textit{CRPhy} \textbf{15} (2014) 329 [arXiv:1403.2967]
%
\bibitem{caprioli14} D. Caprioli et al., \textit{Simulations of Ion Acceleration at Non-relativistic Shocks. II. Magnetic Field Amplification}, \textit{ApJ} \textbf{794} (2014) 46 [arXiv:1401.7679]
%
\bibitem{cardillo15} M. Cardillo, E. Amato $\&$ P. Blasi, \textit{On the cosmic ray spectrum from type II supernovae expanding in their red giant presupernova wind}, \textit{ Astrop. Phys.} \textbf{69} (2015) 1C [arXiv:1503.03001]
%
\bibitem{chevalier82}R. A. Chevalier, \textit{Self-similar solutions for the interaction of stellar ejecta with an external medium}, \textit{ApJ} \textbf{258} (1982) 790
%
\bibitem{chevalier89}R. A. Chevalier et al., \textit{The interaction of supernovae with circumstellar bubbles}, \textit{ApJ} \textbf{344} (1989) 332
%
\bibitem{argo} I. De Mitri et al. (ARGO-YBJ coll.), \textit{Latest results on cosmic ray physics from the ARGO-YBJ experiment}, \textit{NIMPA} \textbf{742} (2014) 2D
%
\bibitem{disciascio14} R. Di Sciascio et al., \textit{Main physics results of the ARGO-YBJ experiment}, \textit{JMPD} \textbf{2330019D} (2013)
%
\bibitem{horandel07} J.R. Horandel et al., \textit{Propagation of super-high-energy cosmic rays in the Galaxy}, \textit{Astrop. Phys.} \textbf{27} (2007) 119 [arXiv:astro-ph/0609490]
%
\bibitem{yac} J. Huang et al., \textit{Primary proton and helium spectra at energy range from 50 TeV to 1015 eV observed with the new Tibet AS core detector array}, \textit{EPJWC} \textbf{5204003H} (2013) 
%
\bibitem{kulsrud69} R. Kulsrud $\&$ W.P. Pearce, \textit{
The Effect of Wave-Particle Interactions on the Propagation of Cosmic Rays}, \textit{ApJ} \textbf{156} (1969) 445
%
\bibitem{lagage83a} O. Lagage $\&$ C.J. Cesarsky, \textit{Cosmic-ray shock acceleration in the presence of self-excited waves}, \textit{A$\&$A}, 118 (1983) 223
%
\bibitem{lagage83b} O. Lagage, O. $\&$ C.J. Cesarsky, \textit{The maximum energy of cosmic rays accelerated by supernova shocks}, \textit{ApJ} \textbf{125} (1983) 249
%
\bibitem{lucek00} S.G. Lucek $\&$ A.R. Bell, \textit{Non-linear amplification of a magnetic field driven by cosmic ray streaming}, \textit{MNRAS} \textbf{314} (2000) 65
%
\bibitem{malkov01} M.A. Malkov $\&$ O. Drury, \textit{Nonlinear theory of diffusive acceleration of particles by shock waves}, \textit{Rep. Prog. Phys.} \textbf{64} (2001) 429
%
\bibitem{ptuskin05} V.S. Ptuskin $\&$ V.N. Zirakashvili, \textit{On the spectrum of high-energy cosmic rays produced by supernova remnants in the presence of strong cosmic-ray streaming instability and wave dissipation}, \textit{A$\&$A} \textbf{429} (2005) 755 [arXiv:astro-ph/0408025]
%
\bibitem{Ptuskin09} V.S. Ptuskin et al., \textit{On leaky-box approximation to GALPROP}, \textit{Astrop. Phys.} \textbf{31} (2009), 284
%
\bibitem{riquelme09} M.A. Riquelme et al., \textit{Nonlinear Study of Bell's Cosmic Ray Current-Driven Instability},\textit{ApJ} \textbf{694} 626 (2009) [arXiv:0810.4565]
%
\bibitem{schure13} K. M. Schure $\&$ A.R. Bell, \textit{Cosmic ray acceleration in young supernova remnants}, \textit{MNRAS} \textbf{435} (2013) 1174 [arXiv:1307.6575]
%
\bibitem{schure14}K.M. Schure $\&$ A.R. Bell, \textit{From cosmic ray source to the Galactic pool}, \textit{MNRAS} \textbf{437} (2014) 2802 [arXiv:1310.7027]
%
\bibitem{skilling75a} J. Skilling , \textit{Cosmic ray streaming. I - Effect of Alfven waves on particles}, \textit{MNRAS} \textbf{172} (1975a) 557
%
\bibitem{skilling75b} J. Skilling,  \textit{Cosmic ray streaming. II - Effect of particles on Alfven waves}, \textit{MNRAS} \textbf{173} (1975b) 245
%
\bibitem{skilling75c} J. Skilling, \textit{Cosmic ray streaming. III - Self-consistent solutions}, \textit{MNRAS} \textbf{173} (1975c) 255
%
\bibitem{truelove99} J.K. Truelove $\&$ C.F. McKee, \textit{}, \textit{ApJ Supp. Series} \textbf{120} (1999) 299
%
\bibitem{TA11} Y. Tsunesada, \textit{Evolution of Nonradiative Supernova Remnants}, \textit{ICRC} \textbf{12} (2011) 67
%
\bibitem{yoon11} Y.S. Yoon et al., \textit{Cosmic-ray Proton and Helium Spectra from the First CREAM Flight}, \textit{ApJ} \textbf{728} (2011) 122-129 [arXiv:1102.2575]
%
\bibitem{wentzel74}D. G. Wentzel, \textit{Cosmic-ray propagation in the Galaxy - Collective effects}, \textit{ARA$\&$A} \textbf{12} (1974) 71
\end{thebibliography}
\end{document}